\RequirePackage{amsthm}

\documentclass[iicol, pdflatex, sn-mathphys-num]{sn-jnl}

\usepackage{graphicx}%
\usepackage{multirow}%
\usepackage{amsmath,amssymb,amsfonts}%
\usepackage{amsthm}%
\usepackage{mathrsfs}%
\usepackage[title]{appendix}%
\usepackage{xcolor}%
\usepackage{textcomp}%
\usepackage{manyfoot}%
\usepackage{booktabs}%
\usepackage{algorithm}%
\usepackage{algorithmicx}%
\usepackage{algpseudocode}%
\usepackage{listings}%

\usepackage{soul} 
\renewcommand{\hl}[1]{#1} 

\usepackage{anyfontsize}
\usepackage[version=4]{mhchem}
\usepackage{multicol}
\usepackage{multirow}
\usepackage[left]{lineno}

\theoremstyle{thmstyleone}%
\theoremstyle{thmstyletwo}%

\theoremstyle{thmstylethree}%

\begin{document}

\title[Article Title]{Multi-messenger dynamic imaging of laser-driven shocks in water using a plasma wakefield accelerator}



\author*[1]{\fnm{M.D.} \sur{Balcazar}}\email{balcazar@umich.edu} 
\author[2]{\fnm{H-E} \sur{Tsai}}\email{haientsai@lbl.gov} 
\author[2]{\fnm{T.} \sur{Ostermayr}}\email{tostermayr@nvidia.com} 
\author[1]{\fnm{P.T.} \sur{Campbell}}\email{campbpt@umich.edu} 
\author[3]{\fnm{F.} \sur{Albert}}\email{albert6@llnl.gov} 
\author[2]{\fnm{Q.} \sur{Chen}}\email{qiangchen@lbl.gov} 
\author[4]{\fnm{C.} \sur{Colgan}}\email{cary.colgan13@imperial.ac.uk} 
\author[5]{\fnm{G.M.} \sur{Dyer}}\email{gilliss@slac.Stanford.edu} 
\author[2]{\fnm{Z.} \sur{Eisentraut}}\email{zgeisentraut@lbl.gov} 
\author[2]{\fnm{E.} \sur{Esarey}}\email{ehesarey@lbl.gov} 
\author[2]{\fnm{C.G.R.} \sur{Geddes}}\email{cgrgeddes@lbl.gov} 
\author[3]{\fnm{E.S.} \sur{Grace}}\email{esgrace@gatech.edu} 
\author[2]{\fnm{B.}\sur{Greenwood}}\email{bgreenwood@lbl.gov} 
\author[2]{\fnm{A.} \sur{Gonsalves}}\email{ajgonsalves@lbl.gov} 
\author[2]{\fnm{S.}\sur{Hakimi}}\email{Sahelh@lbl.gov} 
\author[2]{\fnm{R.} \sur{Jacob}}\email{rjacob@lbl.gov} 
\author[4]{\fnm{B.} \sur{Kettle}}\email{b.kettle@imperial.ac.uk} 
\author[3,6]{\fnm{P.} \sur{King}}\email{pking12@utexas.edu} 
\author[1,7]{\fnm{K.} \sur{Krushelnick}}\email{kmkr@umich.edu} 
\author[3]{\fnm{N.} \sur{Lemos}}\email{candeiaslemo1@llnl.gov} 
\author[4]{\fnm{E.} \sur{Los}}\email{e.los18@imperial.ac.uk} 
\author[1]{\fnm{Y.} \sur{Ma}}\email{yongm@umich.edu} 
\author[4]{\fnm{S.P.D.} \sur{Mangles}}\email{stuart.mangles@imperial.ac.uk} 
\author[1]{\fnm{J.} \sur{Nees}}\email{nees@eecs.umich.edu} 
\author[3,6]{\fnm{I.M.}\sur{Pagano}}\email{ipagano@utexas.edu} 
\author[2]{\fnm{C.} \sur{Schroeder}}\email{cbschroeder@lbl.gov} 
\author[3]{\fnm{R.A.} \sur{Simpson}}\email{razzy@mit.edu} 
\author*[1,7]{\fnm{A.G.R.} \sur{Thomas}}\email{agrt@umich.edu} 
\author[7]{\fnm{M.} \sur{Trantham}}\email{mtrantha@umich.edu} 
\author[2]{\fnm{J.} \sur{van Tilborg}}\email{jvantilborg@lbl.gov} 
\author[2]{\fnm{A.} \sur{Vazquez}}\email{avazquez@lbl.gov} 
\author[7]{\fnm{C.C.} \sur{Kuranz}}\email{ckuranz@umich.edu} 


\affil*[1]{\orgdiv{G\'{e}rard Mourou Center for Ultrafast Optical Science}, \orgname{University of Michigan}, \orgaddress{\street{2200 Bonisteel Blvd.}, \city{Ann Arbor}, \postcode{48109}, \state{Michigan}, \country{United States}}}

\affil[2]{\orgname{Lawrence Berkeley National Laboratory}, \orgaddress{\street{1 Cyclotron Road MS 71-259}, \city{Berkeley}, \postcode{94720}, \state{California}, \country{United States}}}

\affil[3]{\orgdiv{National Ignition Facility and Photon Sciences}, \orgname{Lawrence Livermore National Laboratory}, \orgaddress{\street{7000 East Avenue}, \city{Livermore}, \postcode{94550}, \state{California}, \country{United States}}}

\affil[4]{\orgdiv{John Adams Institute for Accelerator Science}, \orgname{Imperial College London}, \orgaddress{\street{SW7 2AZ}, \city{London}, \state{England}, \country{United Kingdom}}}

\affil[5]{\orgname{SLAC National Accelerator Laboratory}, \orgaddress{\street{2575 Sand Hill Rd}, \city{Menlo Park}, \postcode{94025}, \state{California}, \country{United States}}}

\affil[6]{\orgdiv{Department of Physics}, \orgname{University of Texas at Austin}, \orgaddress{\street{2515 Speedway No. 5.208}, \city{Austin}, \postcode{78712}, \state{Texas}, \country{United States}}}

\affil[7]{\orgdiv{Department of Nuclear Engineering and Radiological Sciences}, \orgname{University of Michigan}, \orgaddress{\street{2355 Bonisteel Blvd.}, \city{Ann Arbor}, \postcode{48109}, \state{Michigan}, \country{United States}}}



\abstract{
\hl{Understanding dense matter hydrodynamics is critical for predicting plasma behavior in environments relevant to laser-driven inertial confinement fusion}. Traditional diagnostic sources \hl{face limitations} in brightness, spatiotemporal resolution, and \hl{inability to detect relevant} electromagnetic fields. \hl{In this work, we} present a dual-probe, multi-messenger laser wakefield accelerator \hl{platform} combining ultrafast X-rays and relativistic electron beams at 1~Hz, to interrogate a free-flowing water target in vacuum, heated by an intense 200~ps laser pulse. \hl{This scheme} enables high-repetition-rate tracking of the interaction evolution \hl{using} both particle types. Betatron X-rays \hl{reveal} a cylindrically symmetric shock compression morphology assisted by low-density vapor, resembling foam-layer-assisted fusion targets. The synchronized electron beam detects time-evolving electromagnetic fields, uncovering \hl{charge separation and} ion species differentiation during plasma expansion --- phenomena not captured by photons \hl{or hydrodynamic} simulations. \hl{We show that combining both probes provides complementary insights spanning kinetic to hydrodynamic regimes, highlighting the need for hybrid physics models to accurately predict fusion-relevant plasma behavior}.
}




\maketitle

\section{Introduction}\label{sec:Introduction}

Inertial confinement fusion (ICF) recently achieved a historical milestone by reaching a target gain of unity \cite{NIF_gain4, NIF_gain5, NIF_gain3, NIF_gain2, NIF_gain1}, reigniting optimism for fusion energy as a transformative and sustainable solution. A key challenge in ICF lies in understanding hydrodynamic instabilities during fuel compression, which are critical for the behavior of the burning plasma. To address this challenge, X-ray radiography has long been an essential tool in high-energy-density physics (HEDP) \cite{Casner, Wan:PRL15, Kuranz:NatCom18}, permitting the diagnosis of plasmas with densities normally exceeding $\rho$ $\geq$ 1 g cm$^{-3}$. High-energy laser facilities commonly irradiate metal, foam, or gas targets with UV laser beams to create X-ray \hl{backlighters}~\cite{WorkmanPCI, Park, fournier2004efficient}. \hl{While these sources coupled with streak cameras can capture many HED-relevant processes, they} suffer from poor brightness due to low conversion efficiency, and limited \hl{spatio-temporal} resolution typically \hl{exceeding $\Delta x > 10$~$\mu$m and $\Delta t > 200$~ps}. 




\hl{In parallel}, electric and magnetic fields are increasingly recognized for playing significant roles in ICF environments~\cite{Li_PRL_2006}. The high temperatures and steep gradients mean that kinetic effects and electromagnetic fields may influence the system dynamics. In this sense, charged particle beam probes have also proven useful in diagnosing these complex interactions \cite{Schaeffer_RMP_2023}. Although previous studies have proposed simultaneous imaging using different particle types~\cite{ostermayr2020laser, orimo2007simultaneous, nishiuchi2008laser}, such as X-rays and protons, they are similarly founded on traditional backlighter-based sources facing many limitations. Some of the challenges include restricted proton energy, spectral modulation, limited simultaneity between the probes, and low temporal resolution. Moreover, these methods typically rely on single-pulse, static configurations, limiting their ability to probe time-dependent dynamic systems. New methods that introduce complementary imaging probes could overcome some of these challenges, offering enhanced spatio-temporal resolution and the integration of field-sensitive diagnostics to study complex plasma interactions. 

As a promising approach laser wakefield acceleration (LWFA) ~\cite{tajima1979laser} has emerged over the past decades as a table-top source of relativistic electrons and X-ray pulses~\cite{mangles2004monoenergetic, geddes2004high, faure2004laser, leemans2006gev, hafz2008stable, wang2013quasi, gonsalves2019petawatt}. The betatron X-rays are ultrafast ~\cite{ta2007demonstration, grigoriadis2021betatron}, have a small source size ~\cite{kneip_nature_phys_2010} enabling high spatial resolution, are brighter than conventional backlighter-based light sources~\cite{albert2014laser}, \hl{and suitable for high-repetition-rate imaging}. Additionally, after propagation of a few cm the photons develop spatial coherence ~\cite{shah2006coherence}, making them \hl{compatible with} phase-contrast X-ray imaging (PCXI) ~\cite{snigirev1995possibilities, Wilkins_Nature_1996, Fourmaux, Vargas} \hl{techniques}. Recent work has demonstrated betatron X-ray capabilities in high-resolution imaging~\cite{Wood18, cole2018high,  Hussein_SR_2019}, X-ray absorption spectroscopy~\cite{mahieu2018probing, kettle2024extended}, among other applications ~\cite{Albert16}. \hl{Moreover,} LWFA electron beams have also been used as probes for laser-matter interactions ~\cite{schumaker2013ultrafast, wan2022direct, zhang2016capturing, zhang2017femtosecond}, aiming to develop an imaging technique that is sensitive to electromagnetic fields and has ultrafast timing resolution. However, previous LWFA studies have been limited exclusively to either a photon or electron probe, without integrating both to achieve a deeper understanding of the system in question.

\hl{In this work we show that utilizing both, high-resolution X-ray photons and coordinated electromagnetic field-sensitive electrons, one can reveal experimental details that would remain hidden from using a single particle type alone}. In a “multi-messenger”-like approach akin to astronomy, one obtains insights about the whole of the interaction, revealing a picture that is greater than the sum of its parts~\cite{MMA_2019}. Moreover, previous LWFA experiments have been restricted to single-shot, solid targets, and many have not been able to exploit the \hl{potential of} high-repetition-rate laser capabilities \hl{in combination with liquid targets} to image time-dependent dynamic systems.

In this study we capture the full dynamic evolution of a laser-heated ablating plasma and shock-compressed water column \hl{using} pump-probe imaging of the interaction with both X-ray and charged-particle beams. X-rays capture the hydrodynamic shock development, while the relativistic electron beam measures the generated \hl{electromagnetic} fields in a non-invasive radiographic scheme. Importantly, both particle sources are correlated in size and synchronized in time. While X-ray or electron beams alone provide an incomplete picture of the interaction, \hl{ther combination offers a powerful and complementary  tool. This approach reveals that detailed knowledge of target conditions along with a holistic view on laser-plasma evolution are needed for accurately modeling the complex plasma dynamics}.

\section{Results}\label{sec:Results}



\begin{figure*}
  \centering
  \includegraphics*[width=\linewidth]{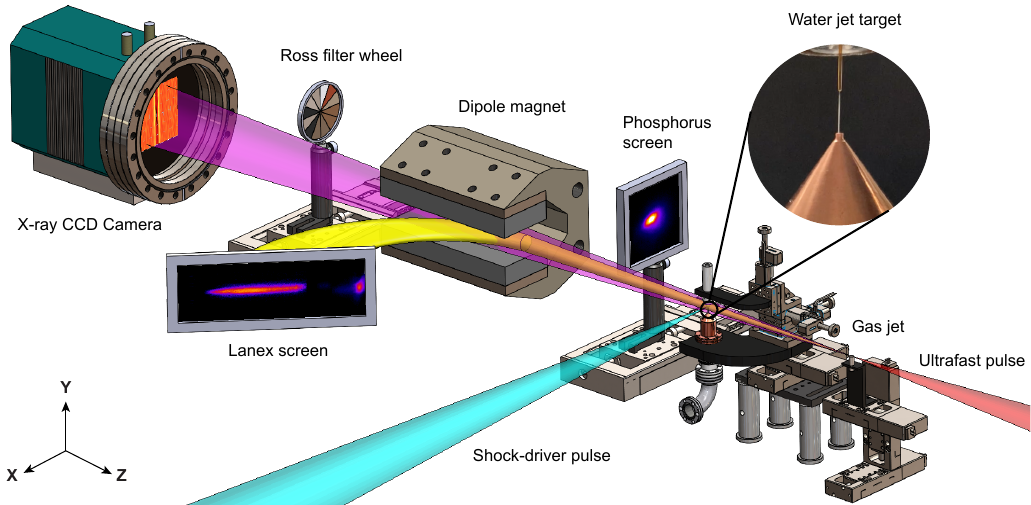}
  \caption{Setup diagram for multi-messenger imaging of laser-driven hydrodynamic shocks in water. The main ultrafast laser pulse is focused on the gas jet driving the plasma wakefield accelerator, generating relativistic electrons and ultrafast X-ray pulses. A secondary long laser pulse is focused into the liquid (water) target creating high-energy-density conditions and driving a hydrodynamic shock. The electron beam profile is recorded with a Phosphorus scintillating screen, and its spectrum is characterized downstream using a magnetic spectrometer which deflects the beam with a dipole magnet into a calibrated scintillating Lanex screen. The betatron X-rays are recorded with an in-vacuum CCD camera, and their spectrum is characterized by using a Ross filter wheel. }
  \label{fig:Figure_1}
\end{figure*}

 The experiments were performed at the Lawrence Berkeley National Laboratory BELLA Center, where the Hundred Terawatt Thomson (HTT) dual beam system was used to drive both the laser wakefield accelerator and the laser target ablator, as shown in Fig.~\ref{fig:Figure_1}. \hl{In combination, a free-flowing liquid (water) target was designed that is capable of providing a bulk plasma (30~$\mu$m diameter stream) in vacuum}. The custom-made water jet (see Sec.~\ref{subsec:MethodsWater}) is replenishable, clean, and suitable for laser-driven high-repetition-rate experiments. \hl{The HTT laser system ($\lambda_{0} = 800$ nm, $1$~Hz repetition rate)} consists on a main LWFA beam line with a linearly-polarized ($1.5 \pm 0.2$ J), ultrafast (pulse duration $40 \pm 5$ fs FWHM, peak power $ 33 \pm 6$~TW) laser pulse focused by an $f$/20 off-axis parabola to a $20 \pm 4$~$\mu$m FWHM spot incident on a gas jet reaching on-target peak intensities of $I =[7 \pm 2] \times 10^{18}$~W cm$^{-2}$. A second ``shock driver'' laser was obtained by splitting the main BELLA HTT beam and subsequently bypassing the compressor, thus generating a long high energy pulse ($200$ ps FWHM, $1.2 \pm 0.2$ J). The secondary pulse was focused on a spot of size $20 \pm 5$~$\mu$m along the $\mathbf{x}$ direction, perpendicular to the main beam axis $\mathbf{z}$, and reaching intensities of $I = [1.4 \pm 0.2] \times10^{14}$~W cm$^{-2}$.  A variable delay stage on one of the beamlines allowed for precise control of timing between the shock-driver pulse with respect to the LWFA probes up to a maximum of $8$ ns within the interaction. Laser pointing fluctuations of $\pm 15~\mu \text{m}$ rms were mitigated by active feedback correction and by setting the focus of the long pulse laser past the water jet, thus employing a spot size larger than the target diameter ($w_{0} \sim 60$~$\mu\text{m}$) at its plane. 

\begin{figure*}
  \centering
  \includegraphics*[width=\linewidth]{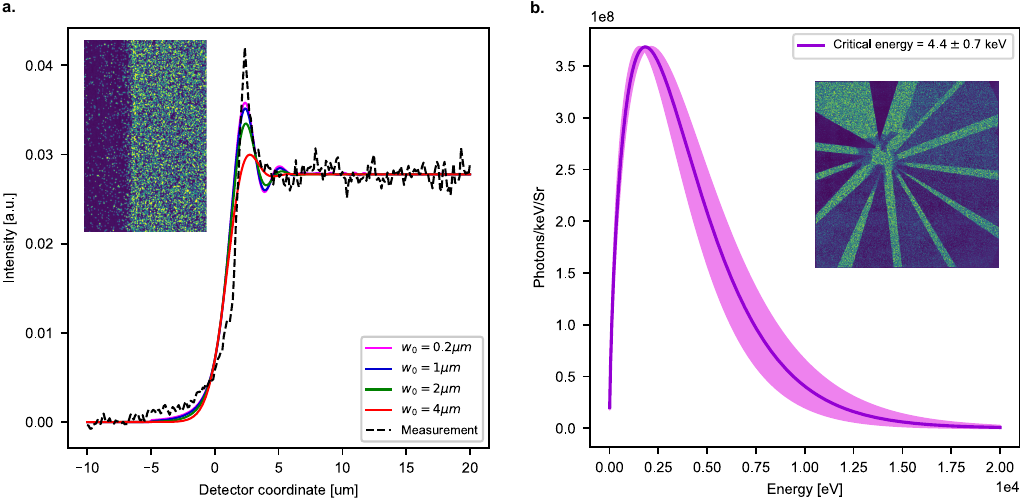}
  \caption{Characterization of betatron X-ray beam probe. \textbf{a} Betatron X-ray source size is measured by imaging a sharp knife-edge and fitting the expected Fresnel diffraction pattern for different source sizes. \textbf{b} Energy spectrum of the betatron X-ray source, recovered by imaging a Ross filter wheel with samples of different materials and thicknesses.}
  \label{fig:Xray-characterize}
\end{figure*}

 Once the short laser pulse travels through the \hl{gas} target (see Sec.~\ref{subsec:MethodsGasJet}) the ponderomotive force expels electrons away from high-field intensity regions leaving ion cavities and launching plasma waves on its wake. Due to the high-field gradients following the pulse, trapped electrons can be accelerated to relativistic velocities during the interaction. The oscillation of these electrons inside the ion ``bubbles'' in turn generates betatron X-ray pulses of ultrafast duration. Therefore, a laser wakefield accelerator generates synchronized X-rays and relativistic electrons, providing a unique \hl{tool for imaging} unlike others where only one type of particle is available. The electron beam probe having an average energy of \hl{$146 \pm 7$} MeV (Fig.~\ref{fig:ebeam-spectrum}) was characterized using scintillating screens and a calibrated dipole magnet downstream. The betatron X-rays were recorded using a cooled, in-vacuum, CCD camera (PI-MTE) and later characterized using a sharp knife-edge and Ross pair filter wheel to have \hl{an} estimated source size to be in the order of $1$~$\mu$m in Fig.~\ref{fig:Xray-characterize} mean critical energy of $4.4 \pm 0.7$ keV (Fig.~\ref{fig:Xray-characterize}). The characterization results of both probes, along with the experimental geometry chosen for multi-messenger imaging are discussed in Sec.~\ref{sec:Methods}. 

Initially, when the long pulse laser interacts with the water, it deposits its energy over some range of densities below critical density, in the corona of the pre-formed plasma. This occurs mostly through inverse bremsstrahlung, but in addition through potential hot electron generation mechanisms \cite{Kruer_book}. Early preheating of the interior may also occur via laser shine-through effects~\cite{bradley1991early}. Following laser energy deposition, electrons carry the heat to higher densities above critical density. Consequently, the dense liquid surface is quickly heated, leading to its ablation and subsequent expansion into vacuum. In response to the large pressures generated at the ablation surface from the expansion of the material a hydrodynamic shock wave is launched into the liquid target. 

For low-Z laser-irradiated targets, reduced opacity minimizes the role of thermal radiation emission in the dynamics, \hl{ and the low atomic number significantly reduces bremsstrahlung background from electron–target interactions compared to high-Z materials. However,} in liquid water ($\rho = 1$~g cm$^{-3}$), partial ionization increases opacity, making radiative losses more significant. At these ICF-relevant densities the evolution of the target is complex, involving hot-electron production, electron heat transport, radiation processes, shock-wave generation, and hydrodynamic expansion – all occurring within fractions of a nanosecond under typical experimental conditions. Nanosecond timescales fall within a regime where the complete physics of the interaction is hardly accessible to any single contemporary simulation. Our setup is then ideally suited for experimentally studying such complex dynamic systems, offering ultrafast and sub-micron resolution, nanosecond delay range, and dual probes sensitive to variation in both density and electromagnetic fields.


\begin{figure*}[!tbh]
  \centering
  \includegraphics*[width=\linewidth]{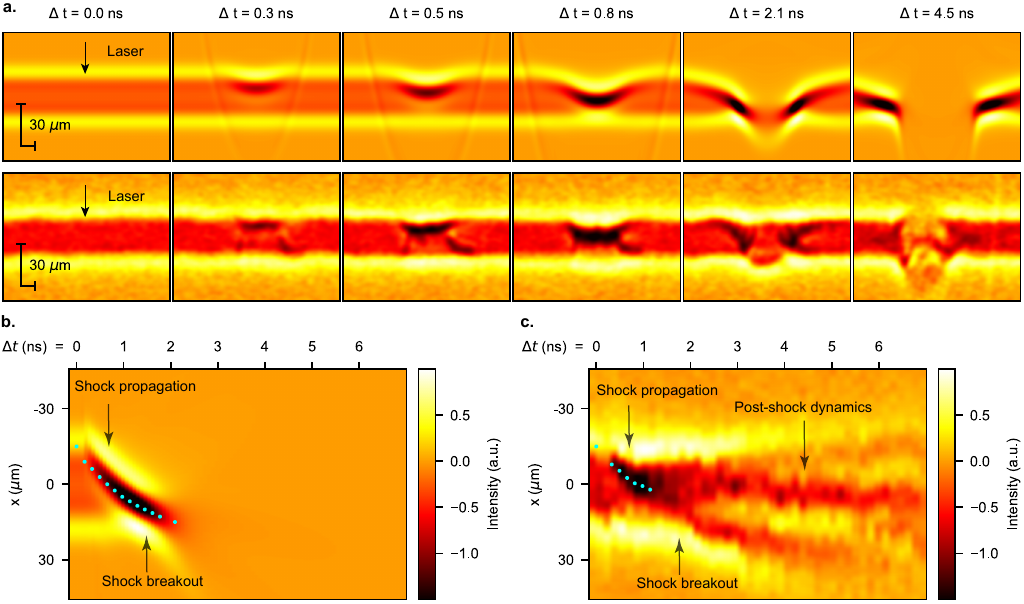}
  \caption{\hl{Comparison between} betatron X-ray imaging \hl{and 3D FLASH hydrodynamic  simulations} of laser-driven shocks in water. \textbf{a} Contrast between experimental betatron X-ray images and synthetic phase-contrast X-ray image (see Sec~\ref{subsubsec:ImageFresnel}) \hl{at different time delays}. \textbf{b} and \textbf{c} \hl{Lineouts averaged over 80 pixel rows} taken at the center of \hl{simulation and experimental} images, and stacked together horizontally to obtain a composite picture of the full delay of the interaction. \hl{Cyan dots indicate tracking of shock position.} }
  \label{fig:Figure_2}
\end{figure*}



The first result, displayed in Fig.~\ref{fig:Figure_2} presents a time-series comparison between synthetically generated phase-contrast X-ray images and experimentally obtained betatron X-ray images. The laser-driven shock displays very similar features in both simulation and experiment. \hl{The features observed in Fig.}~\ref{fig:Figure_2}a \hl{include bright phase-contrast enhancement at the edges of the target, as well as a strong, dark, bow-shaped shock structure that grows in $\mathbf{y}$-direction and propagates forward in $\mathbf{x}$-direction. Smaller precursor signals ahead of the main shock are also appreciable and are explored further in the results.} The evolution of the laser-water interaction was predicted using FLASH~\cite{fryxell2000flash}, a radiation hydrodynamic simulation code. Projected density maps obtained from FLASH were post-processed with a Fresnel-Kirchoff-based algorithm (Sec.~\ref{subsubsec:ImageFresnel}) to produce the synthetic phase-contrast X-ray patterns as shown in Fig.~\ref{fig:Figure_2}a. 

\begin{figure}[!tbh]
  \centering
  \includegraphics*[width=\linewidth]{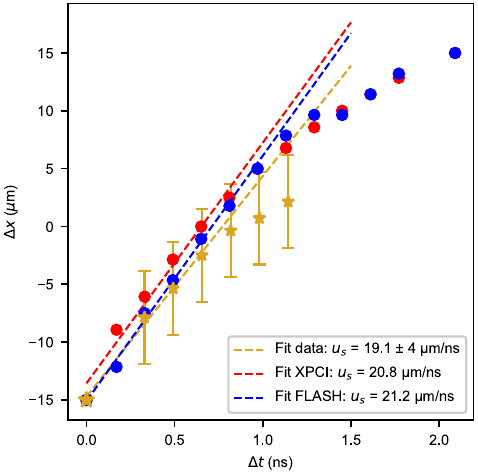}
  \caption{ \hl{Hydrodynamic shock velocity analysis comparing simulation and experiment.}. The velocity measurement tracks the point of highest density in FLASH, and the minimum intensity feature in phase-contrast X-ray images, both for synthetic and experimental cases.}
  \label{fig:Figure_3}
\end{figure}

To analyze these results, average lineouts were taken at the center of each image in the time-series. The lineouts were then stacked together to create a composite image of the full interaction, shown in \hl{Fig.}~\ref{fig:Figure_2}b \hl{and} Fig.~\ref{fig:Figure_2}c. \hl{These} aggregate images reveal the main shock breakout through the rear of the target around $\Delta t = 1.5 - 2$~ns. The lineouts were subsequently analyzed to compare the shock velocity between simulation and experiment, which are comparable in magnitude as demonstrated in the analysis of Fig.~\ref{fig:Figure_3}.   

Measuring shock velocity is particularly important because it can provide information regarding the thermodynamic state of the material. \hl{By} taking the frame of reference where the shock is at rest, one can define an upstream shock Mach number , $M_{u} = u_{s} \sqrt{ {\rho_{1} }/{ ( \gamma}p_{1})}$. Here $u_{s}$ represents the shock velocity, $\rho_1$ denotes the unshocked fluid density, $p_{1}$ is the corresponding initial pressure, and the polytropic index value is assumed to be $\gamma = 5/3$. \hl{Such index is typical for fully ionized, weakly coupled HED systems where radiative effects are minimal}~\cite{drake2018}. For the present experimental parameters $M_{u} \sim 10^{4}$, which is a clear indication of the strong-shock regime. 

In order to calculate useful parameters for ICF \hl{environments}, one can utilize the well-known Rankine-Hugoniot jump conditions and obtain predictions for the post-shock density and pressure in terms of $M_u$ as,
\begin{equation}
    \frac{\rho_{2}}{\rho_{1}} = \frac{M_{u}^{2} (\gamma + 1)}{M_{u}^{2}(\gamma - 1) + 2}\;, 
    \label{eq:jumprho}
\end{equation}
and
\begin{equation}
    \frac{p_{2}}{p_{1}} = \frac{2 \gamma M_{u}^{2} - (\gamma - 1) }{(\gamma + 1)}\;.
\end{equation}
The uncertainty in $M_{u}$ can be pretty significant, however note that in the limit when $M_{u} \gg 1$, the physical limit for density jump in a polytropic gas tends towards $\rho_{2}/\rho_{1} = (\gamma + 1)/(\gamma - 1) \sim 4$, while the pressure grows infinitely, and is strongly dependent on the shock velocity measured experimentally. For the analysis shown in Fig.~\ref{fig:Figure_3}, utilizing \hl{$u_{s} = 20$~$ \mu\text{m}/\text{ns}$}, the resulting post-shock \hl{pressure on target is $p_{2} = 3 - 4$ {Mbar}, comparable in magnitude to that at the Earth's core}. \hl{This pressure scales with laser intensity as $p_{2} \approx 6.1 I_{14}^{2/3} \lambda_{u}^{-2/3}$ Mbar. Where $I_{14}$ is the intensity in units of $10^{14}$~W cm$^{-2}$ and $\lambda_{u}$ is the laser wavelength in micrometers} ~\cite{drake2018}.

Moreover, one can utilize the calculated pressure $p_{2}$ and the measured shock velocity $u_{s}$ to obtain a useful prediction for the ion temperature. By taking $p_{2} = \left(Z_{2} + 1\right) k_{B} T_{2} \rho_{2}/\left(A m_{p} \right)$ and taking the strong shock limit one finds
\begin{equation}
    k_{B} T_{i,2} = \frac{A m_{p}}{\left( 1 + Z_{2} \right)} u_{s}^{2} \frac{2 \left(\gamma - 1 \right)}{\left( \gamma + 1 \right)^{2}} \;.
\label{eq:T2}
\end{equation}
 Assuming the electrons are non-degenerate, they fully equilibrate with the ions, and ignoring Coulomb modifications to the pressure, the immediate postshock temperature of the ions before they equilibrate with the electrons can be found by setting $Z_{2} = 0$, as the average ionization of the post-shock state. Then Eq.~\ref{eq:T2} results in an ion temperature of \hl{$T_{i,2} = 4.7$~{eV}}.

\begin{figure}[!tbh]
  \centering
  \includegraphics*[width=\linewidth]{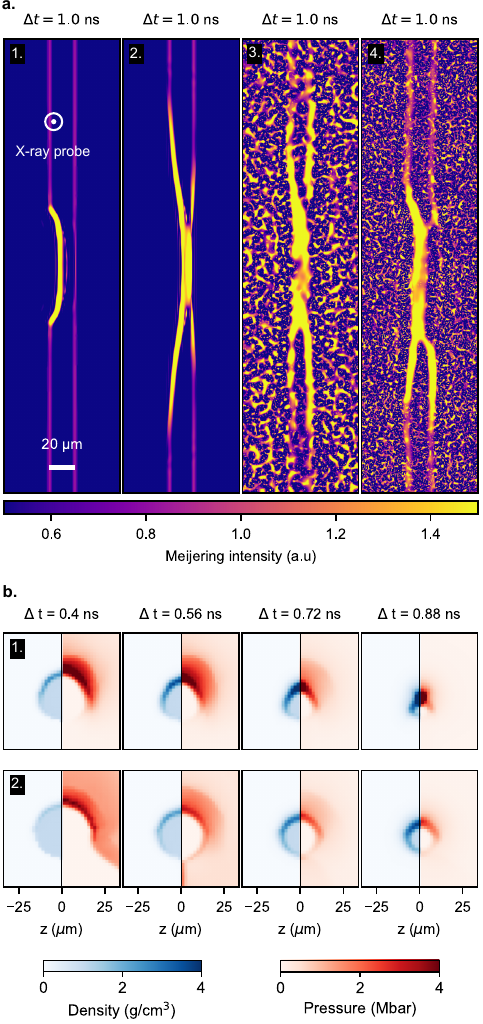}
  \caption{Demonstration of cylindrically symmetric shock compression morphology assisted by low-density vapor. \textbf{a} Comparison panels with Meijering filter (see Sec~\ref{subsubsec:ImageMeijering}) at $\Delta t\sim 1.0$~ns. Panel (1) is a $\mathbf{x}$-$\mathbf{y}$ 2D slice from 3D FLASH simulation with no vapor, panel (2) is a $\mathbf{x}$-$\mathbf{y}$ 2D slice from 3D FLASH simulation with surrounding vapor profile, panel (3) is a ten-shot-averaged image taken with betatron X-rays, 4) single-shot image taken with betatron X-rays. \textbf{b} $\mathbf{z}$-$\mathbf{x}$ 2D slices from 3D FLASH simulation comparing density and pressure maps as a function of time for two cases: panel (1) water target without surrounding vapor and panel (2) water target with surrounding vapor profile.}
  \label{fig:Figure_4}
\end{figure}

Notably, although early time agreement with simulations was good regarding the shock propagation velocity, \hl{the simulated shock structures did not fully match the experimental measurements at intermediate times}. In Fig.~\ref{fig:Figure_4}a\hl{, panel (1)} represents a simple 3D (or standard 2D), FLASH simulation of a cylindrical water target heated by a laser pulse. Interestingly, ten-shot averaged \hl{panel (3)}  and single-shot \hl{panel (4)} show X-ray images of the shocked target at near maximum compression at $\Delta t\sim 1.0$~ns. These display an apparent rear-driven shock structure, that is different from \hl{conventional} 2D, and simple 3D, FLASH simulations. 

Analysis of the plasma conditions indicate that it is unlikely to observe a strong shock reflection at the rear interface. Instead, a much better agreement with the structure produced experimentally was achieved in Fig.~\ref{fig:Figure_4}a panel (2), by incorporating a low-density layer surrounding the target, approximating the vapor expected from water evaporation in vacuum (see Sec.\ref{subsec:MethodsFLASH} for simulation details). Including the low-density layer in the simulations enhances thermal transport around the water column, resulting in more uniform heating of the jet’s exterior and producing a cylindrical shock \cite{laso2024cylindrical} and symmetric compression morphology closely matching the experimental observations. 


To illustrate this effect, Fig.~\ref{fig:Figure_4}b displays 2D slices from 3D simulations of the density and pressure profiles taken at the midpoint, $\mathbf{z}-\mathbf{x}$ plane, of the water cylinder. \hl{Panel (1)} shows a simple water target with no surrounding vapor, in which the laser generated shock is strong and predominantly one-sided. In contrast, if a vapor density profile falling as $\propto 1/r$ from the target surface is introduced in \hl{panel (2)} the magnitude of the pressure and density of the shock are decreased. More importantly for our discussion is the difference in compression morphology, where the \hl{partially ionized} vapor blanket provides a \hl{low-opacity, low-density medium in which electrons are able to transport energy more efficiently} around the surface of the target. \hl{This leads to more uniform pressure and density profiles, resulting in a cylindrically symmetric shock compression morphology}.

\begin{figure*}
  \centering
  \includegraphics*[width=0.9\linewidth]{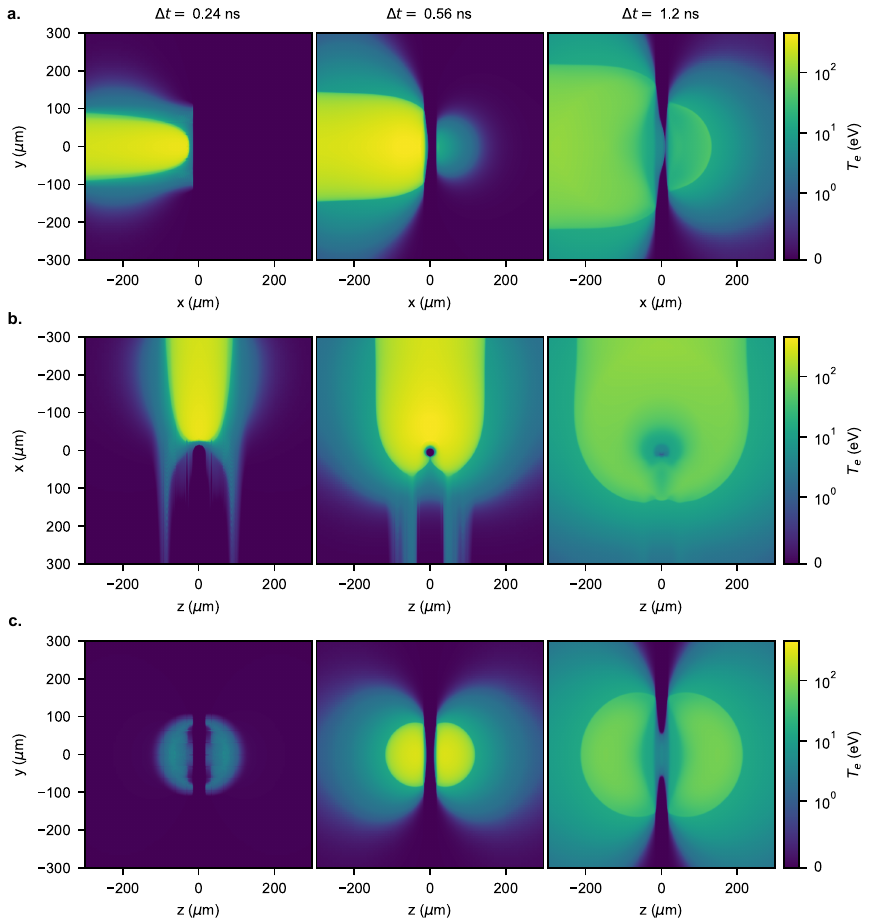}
  \caption{Electron temperature $T_{e}$ evolution during laser-water interaction. \hl{2D slices are captured from 3D FLASH simulations}: \textbf{a} x-y plane, \textbf{b} z-x plane, and \textbf{c} z-y plane.}
  \label{fig:ED-flash-tele}
\end{figure*}

To obtain a deeper understanding of this phenomenon, consider that the laser energy at the peak of the pulse is deposited in ionized water electrons below critical density $n_{c}$. This creates a high-temperature region of order $\sim1$~keV in the simulations (Fig.~\ref{fig:ED-flash-tele}), from which electrons transport heat above $n_{c}$ into the dense matter~\cite{delettrez1986thermal}. Free electrons in the low-density corona region are not able to fully escape from the electrostatic forces imposed by the ions, but are free to move around the surface of the target traveling a significant distance before depositing their energy. For instance, let us compare the mean-free-path of a 1~keV thermal electron carrying heat in the plasma region below $n \leq n_{c}$ to that of an electron in the dense solid $n \approx n_{e0}$. For the low density region with $n_{c}\sim 10^{21} \text{cm}^{-3}$ we obtain $\lambda_{mfp} \sim 20$ $\mu\text{m}$, \hl{in the order of the diameter of the stream}. By contrast, for the high-density region $n_{0}\sim 10^{23}$ $\text{cm}^{-3}$ the mean-free-path is $\lambda_{mfp} \sim 0.2$ $\mu\text{m}$. As a result, the heat flow within the target would be localized, while thermal transport in the vapor layer would be highly nonlocal.


\begin{figure*}[!tbh]
  \centering
  \includegraphics*[width=\linewidth]{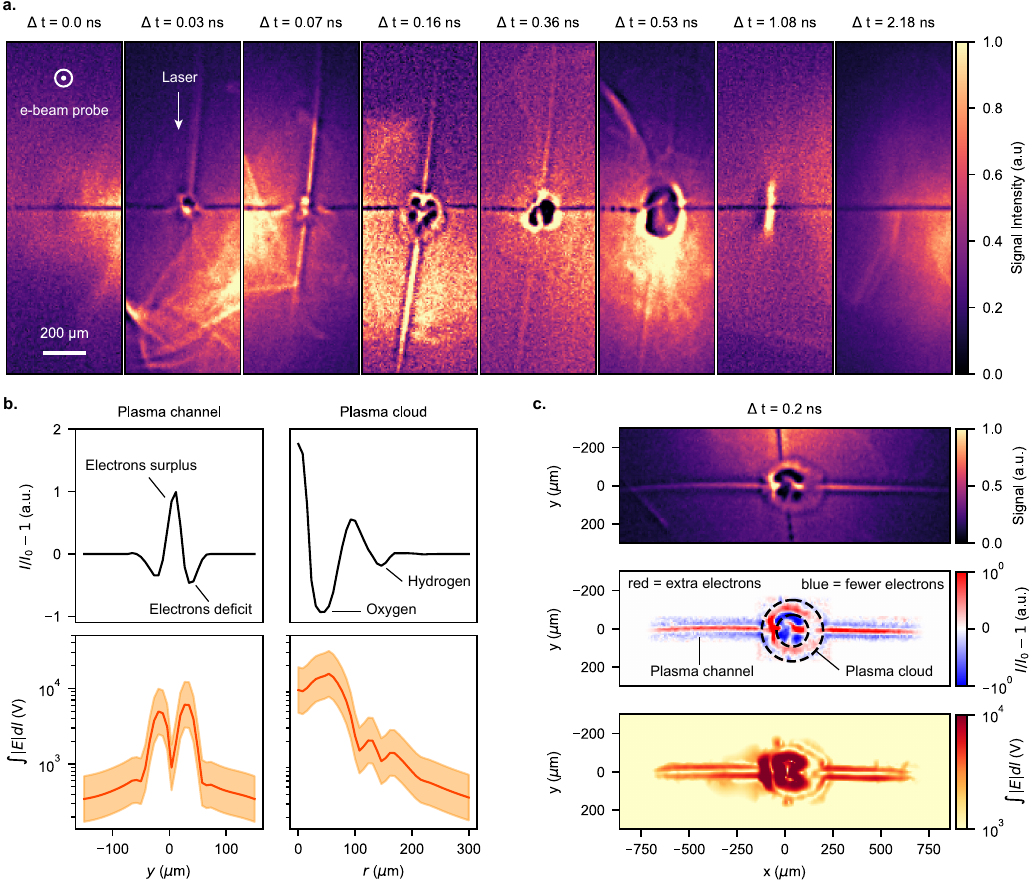}
  \caption{Dynamic probing of \hl{time-evolving} electromagnetic fields in water using the LWFA relativistic electron beam \hl{probe}. \textbf{a} Time-series of electron beam profile disturbed by electromagnetic fields around the laser-water interaction region \hl{recorded on Phosphorus screen}. \textbf{b} Integrated line profiles for the plasma channel and plasma cloud \hl{features} for both the difference image $I/I_{0} - 1$, and recovered electric fields $\int |E| dl$ \hl{including error bands accounting for the uncertainty in their magnitude from chromatic effects}. \textbf{c} Illustration of electric field recovery from electron beam radiographic images (see Sec~\ref{subsubsec:ImageFields}). \hl{The panel includes the radiographic} normalized image \hl{$I$, the difference image $I/I_{0} - 1$}, and recovered field magnitude $\int |E| dl$. Dashed circles highlight the two distinct features, an Oxygen plasma (inner circle) and a Hydrogen plasma (outer circle).}
  \label{fig:Figure_5}
\end{figure*}

The experimental observations in ~\ref{fig:Figure_4}a panel (3) and panel (4) have led us to better understand the compression morphology of the target, and the importance of accurate modeling the experimental conditions. To \hl{to support this interpretation with observational} evidence, we now make use of the relativistic electron beam probe generated by the laser wakefield accelerator. The nonlocality of the heat transport, and the presence of energetic electrons around the interaction region result in the production of electric (through the strong pressure gradients) and magnetic (through current flows) fields in the plasma surrounding the water target. These field-generating mechanisms are well known but are often not included in fluid-based hydrodynamic codes, as electric fields are typically precluded by enforced quasineutrality. Unlike the X-ray probe, the LWFA electron beam will be sensitive to path-integrated deflections caused by electric and magnetic fields along its trajectory. \hl{These perturbations are captured downstream on the profile imager, forming an image of the interaction.}


\hl{Using this approach}, the electron beam probe was used to capture the \hl{evolution} between the long pulse laser and the water stream\hl{, as shown} in Fig.~\ref{fig:Figure_5}a. Bright features displayed in the scintillator screen indicate accumulation of probe electrons from focusing fields, while dark features would typically represent absense of electrons from defocusing fields. The first noticeable feature in the time-series is a bright channel in the laser propagation direction, which persists for approximately the pulse duration $200-300$~ps and focuses the electron probe. The fields are likely to be electric arising from the pressure gradient caused by the long pulse laser ionization and heating. Notably, the observation of such early-time ionization channel across both sides of the water jet further supports the idea of nonlocal heat transport, hot electron generation, and a symmetrically-ablated target observed with the X-rays in Fig.~\ref{fig:Figure_4}. Although the channel forming on both sides of the liquid jet might suggest laser transmission through the water, it is important to recall that the long-pulse laser beam is larger than the water column diameter, allowing a significant portion of the laser energy to bypass the target.

\begin{figure*}
  \centering
  \includegraphics*[width=\linewidth]{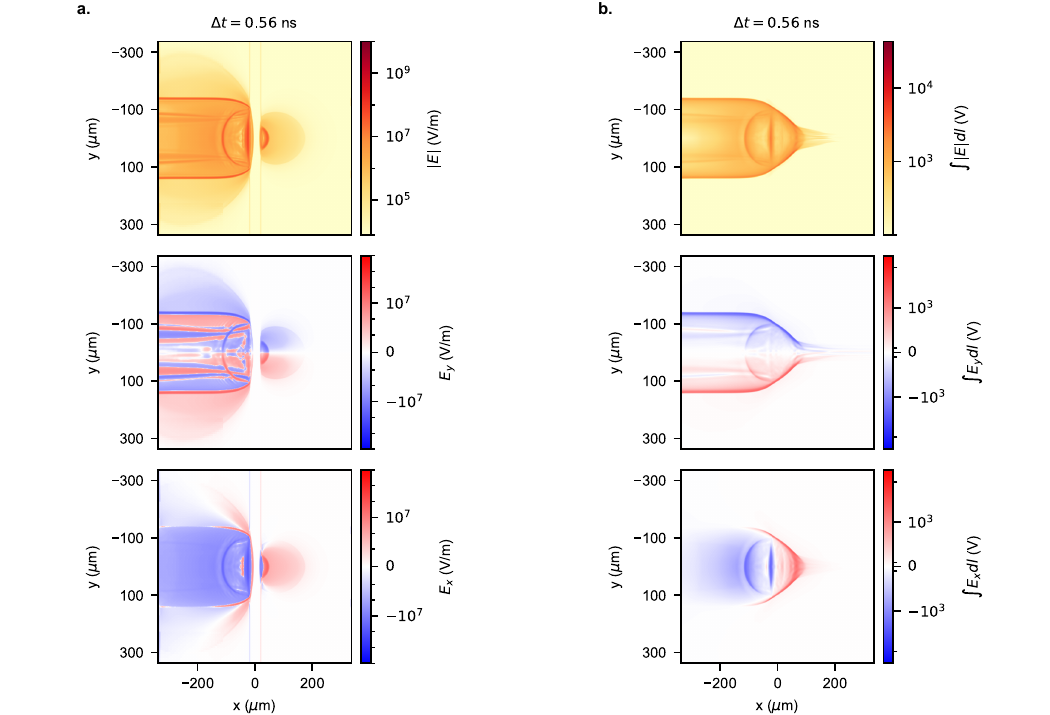}
  \caption{\hl{Illustration of derived electric fields from 3D FLASH simulations. The fields are obtained following the equation $E \sim \nabla P / e n_{e}$, where $P$ is the pressure field and $n_{e}$ is the electron number density field outputs from FLASH. \textbf{a} 2D slice of electric fields displaying  $E_x$, $E_y$ and magnitude $|E|$. \textbf{b} Projected electric fields across the spatial domain including $E_x$, $E_y$ and magnitude $|E|$.} }
  \label{fig:ED-FLASH-fields}
\end{figure*}

The second prominent feature observed with the electrons is a dark plasma cloud expanding from the center, indicating the presence of complex, strong electromagnetic fields. While this structure might initially be interpreted as defocusing fields, the absence of a plausible physical mechanism (see Fig.~\ref{fig:ED-FLASH-fields}) suggests the fields are most likely over-focusing the probe. Moreover, the sharp bright circular ring surrounding the plasma cloud indicate the presence of caustics in the image. \hl{Although caustics complicate quantitatively analysis of the fields, lower bounds on their strengths and key features of their topology may still be extracted using a field recovery method (Sec.}~\ref{subsubsec:ImageFields}). From Fig.~\ref{fig:Figure_5}c, we infer integrated fields on the order of $\int Edl\sim 10^4$~V or $\int Bdl\sim 10^{-4}$~T$\cdot$m. In the case of electric fields, a simple estimate using $E\sim\nabla P/en_e\sim \nabla (k_BT_e/e)$ would imply hot electrons of order keV temperature. 

More importantly, \hl{the} morphological understanding of the fields \hl{is} acquired from Fig.~\ref{fig:Figure_5}, where different plasma species can be identified from the radiographic images. First, a dark inner cloud bounded by a bright ring, likely caused by strong overfocusing fields from an Oxygen ion plasma. Secondly, a fainter dark outer ring, which may be attributed to a Hydrogen ion (protons) plasma marking the boundary of its expansion into vacuum. At the plasma edge, a sheath field forms due to the separation of electrons and protons by an amount of around the local Debye length. This sheath field balances the thermal motion and exerts a positive force on the electrons in the negative radial direction, focusing (or overfocusing) the probe. \hl{The timescale of expansion indicates that the dark outer ring must originate from an expanding proton plasma defocusing the probe, as it expands too slowly to be exclusively attributed to defocusing electrons given the working temperatures}.

\begin{figure}[!tbh]
  \centering
  \includegraphics*[width=\linewidth]{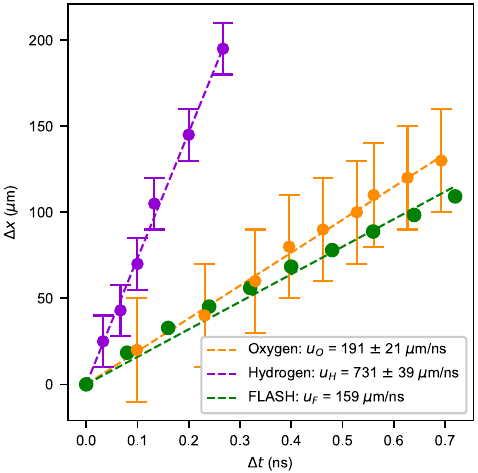}
  \caption{Analysis of laser-ablated plasma expansion velocity and comparison with simulation. For FLASH simulation the measurement tracks the density edge position of the expanding plasma plume. For electron beam radiographic images the measurement tracks annular features expanding in time, where two distinct plasma species components are identified: (1) Hydrogen ions plasma and (2) Oxygen ions plasma.}
  \label{fig:Figure_6}
\end{figure}

Within this framework, the expansion velocity analysis in Fig.~\ref{fig:Figure_6} presents both a slower Oxygen plasma (inner cloud) with velocity of $u_{O} = 191 \pm 7$~$\mu$m/ns that, to within error, agrees with the expansion of the plasma in FLASH simulations, $u_{F} = 159$~$\mu$m/ns. As well as a second Hydrogen plasma species (outer dark ring) with a much faster characteristic expansion speed of $u_{H} = 731 \pm 74$~$\mu$m/ns. The speed of this plasma boundary is consistent with protons moving with the same average energy as the slower expanding Oxygen ions. Protons are expected to be accelerated from the surface of a hot plasma \cite{Wilks_POP_2001}, however, this feature is not present in the single-fluid FLASH \hl{simulations.

Hence,} the relativistic electron beam probe is able to provide evidence for an expanding hot plasma surrounding the water column not visible by the X-rays, as well as capture the evolution of different ion species not present in the simulation. More importantly, the dual-sided near-isotropic ablating plasma, observed with the electron beam probe, would support the model of a cylindrically symmetric compression shock morphology observed with the X-ray\hl{s}.  

\hl{The absence of hot electron populations in FLASH simulations is particularly relevant in light of our observations. Hot electrons generated during laser–plasma interactions can travel significant distances, driving non-local heat transport and depositing energy into the surrounding vapor layer. These findings underscore the need for future simulations that incorporate kinetic effects alongside radiation hydrodynamics, as well as advanced diagnostic systems in HED experiments.}


\section{Discussion}\label{sec:Discussion}

We have shown that combining information in tandem from both an ultrafast X-ray probe and a relativistic electron beam probe from \hl{LWFA enables a multi-messenger technique that reveals insights inaccesible to either probe alone. This approach} provides a more comprehensive understanding of the physics involved \hl{in} high intensity laser \hl{interactions} with dense matter. While betatron X-ray imaging captures laser-driven shock hydrodynamics with submicron resolution, electron beam radiography offers a complementary perspective on electromagnetic field \hl{time} evolution. LWFA, in contrast to other radiation sources, generates simultaneous electron beam and X-ray beams with good beam properties, \hl{intrinsic femtosecond} time resolution, excellent absolute timing and synchronization with themselves as well as other lasers. These advantages make it a unique tool that is hardly realized by synchrotron or X-ray free-electron laser facilities, where only one type of particle probe is available for experiments.


\hl{While the simulated early shock propagation velocity reasonably matched experimental measurements, discrepencies} in the shock compression morphology \hl{were observed when compared to 2D and simple 3D fluid simulations. In these simplified cases, simulations predominantly showed} single-sided planar ablation \hl{from the long pulse laser}. In \hl{contrast}, the multi-messenger probe \hl{revealed} a more \hl{holistic} picture of the laser-water interaction, demonstrating a cylindrically symmetric shock compression morphology. By refining the simulation model to more precisely account for initial conditions of the target, such as vacuum-induced water evaporation, a more accurate representation of the interaction was obtained. 

This vapor-assisted cylindrically-symmetric compression phenomenon can be compared, in some ways, to low-density foam-layer-assisted inertial confinement fusion targets~\cite{moore2020experimental, bhandarkar2018fabrication}, concerning advanced hohlraum designs with reduced wall-motion in ICF. The ablative expansion morphology of laser-heated materials is a crucial factor in optimizing hohlraum cavities for \hl{ICF}, where low-density foam liners help manage the expansion of the heated hohlraum walls and decrease the development of instabilities. \hl{Previous experiments have demonstrated effectiveness in mitigating wall motion that interferes with a fully spherically-symmetric compression of the D-T fusion target through liners with densities in the order of $\rho = 0.01-0.04$~g cm$^{-3}$ . In an analogous sense, the low density water vapor layer with $\rho_0 = 0.01$~g cm$^{-3}$ in this experiment serves to assist the single laser ablator to compress the target with a cylindrically-symmetric morphology.} \hl{At later times, the X-ray probe further captures complex dynamics consistent with the onset of post-shock plasma instabilities, as expected by the experimental conditions. Some of these features are appreciated in Fig.}~\ref{fig:Figure_2}c \hl{at $t > 3$ ns, as signals not discerned in the simulation model. A detailed analysis of these features is underway and will be presented in future work.} 

\hl{Utilizing} the electron beam probe further revealed quasi isotropic expanding plasma \hl{fields} on both sides of the target, capable of \hl{identifying} distinct plasma  species expanding at characteristic thermal velocities --- features \hl{not accessible by photon-based imaging}. These observations not only advance our understanding of laboratory plasma physics but also underscore discrepancies between radiation hydrodynamic simulations and real laser-plasma interactions, where complex phenomena such as ion species differentiation and \hl{strong} electromagnetic fields are evident \hl{from femtosecond to nanosecond timescales}.

\hl{All measurements were repeated across multiple shots (typically 5–10 per delay time) under the same conditions. Despite the fluid nature of the target, key features such as shock evolution, compression morphology, and field topology were reasonably reproduced across the dataset.} \hl{That said, the dual-probe setup does involves certain trade-offs, including spatial constraints on sample and detector placement, characterization of each probe, as well as added complexity in target engineering.}

\hl{While this configuration may not directly replicate full-scale direct-drive ICF experiments, the purpose of this work is to demonstrate the diagnostic power of LWFA based multi-messenger probes and to set the path forward for more sophisticated, high-repetition-rate platforms using liquid targets. These insights may also motivate the design of future smaller-scale HED science experiments, which could one day be adapted to large-scale fusion facilities such as the National Ignition Facility, Laser MegaJoule, or OMEGA.}
 


\section{Methods}\label{sec:Methods}


\subsection{\label{subsec:MethodsXrayImaging} X-ray beam imaging}

The experimental geometry was carefully designed to enable a propagation-based phase-contrast X-ray imaging configuration. In order to properly choose a source-to-object distance $R_{0}$ and object-to-detector distance $R_{1}$ , the coherence length $L_{\perp} = R_{0}/k \sigma$ was estimated for a collection of uncorrelated emitters of wavenumber $k$ and source size $\sigma$ \cite{kneip_nature_phys_2010}. To resolve features of order $\sigma$ at the image plane the coherence length must exceed the source size ($L_{\perp} > \sigma$). Rearranging, the source-to-object distance can be estimated as: $R_{0} > k \sigma^{2}$, which in terms of the photon energy in units of keV and microns is
\begin{equation}
\centering
    R_{0} [m] > 5.1 \times 10^{-3} \left( \sigma [\mu \text{m}] \right)^{2} \left(\hbar \omega [\text{keV}] \right)\;.
\end{equation}
For $5$~keV X-rays and a $\sigma \approx 2\mu$m source size then $R_{0} > 10$~cm ensures that the coherence length is larger than the source size. 

The broadband polychromatic spectrum of a betatron source supports propagation-based phase contrast since, to first order in the paraxial approximation, the phase-contrast pattern is independent of wavelength when the effective Fresnel number is larger than unity $N_{F,eff} > 1$, where
\begin{equation}
\begin{split}
   N_{F,eff} = &\left( R \mathbf{\lambda} \right)^{-1} \frac{M}{M - 1} \sigma_{\text{obj}} \\
   &\times \sqrt{M^{2} \sigma_{\text{obj}}^{2} + \left(M - 1\right)^{2} \sigma^{2} + \text{PSF}^{2} }
\end{split}
\end{equation}
Here \hl{$\mathbf{\lambda}$} is the weighted mean wavelength, $R = R_{0} + R_{1}$, $M = \left(R_{0} + R_{1} \right) / R_{0}$ is the magnification, $\sigma_{\text{obj}}$ is the size of the smallest object feature, and PSF represents the detector's point-spread function \hl{as described in Ref.}~\cite{barbato2019quantitative}. Therefore, the condition $N_{F,eff} >1$ holds true given our experimental setup configuration with source-to-object distance \hl{$R_{0} = 16$~cm}, object-to-detector distance $R_{1} = 580$~cm, and beam divergence of a few tens of mrad. 


\subsection{\label{subsec:MethodsXrayCharacterize} X-ray beam characterization}

The X-ray radiation spectrum was characterized using a Ross filter wheel, as described in Ref.~\cite{PMKing2019}. The filter wheel consisted of wedges made of different materials and thicknesses to selectively attenuate parts of the spectrum. By analyzing the transmitted intensity map and fitting a synchrotron-like spectrum~\cite{esarey2002synchrotron}, the critical energy of the X-ray beam was determined to be  $E_{\text{crit}} = 4.4 \pm 0.7$~keV.

The size of the X-ray source was estimated using a sharp “knife-edge” placed in the beam path. The intensity profile of the transmitted beam was measured and fitted to the expected Fresnel diffraction pattern produced by a Gaussian source interacting with a half-plane. This analysis yielded an upper bound for the X-ray source size of  $\sigma \leq 1$~$\mu$m.

\hl{To obtain a nominal background signal, we recorded images without gas or water flow and averaged over 10 shots to improve the signal-to-noise ratio. The background was then subtracted from each signal image on a shot-by-shot basis. To protect the in-vacuum CCD camera, and further reduce background signals, a filter assembly was placed downstream from the detector. This consisted of a 60 $\mu$m Aluminum foil, a 25 $\mu$m Kapton film, and a 36 $\mu$m Mylar layer, which together served to attenuate residual laser light, suppress plasma self-emission, and shield the detector from debris.}


\subsection{\label{subsec:MethodsEbeamImaging} Electron beam imaging}

The geometry for electron-beam imaging shared the same source-to-object distance as the X-ray imaging setup $R_{0}^{*} = R_{0}$ but used a significantly shorter object-to-detector distance $R_{1}^{*} = 141$~cm. This configuration was selected to capture the full electron beam profile on a Phosphorus scintillating screen located downstream in the chamber, prior to the dipole magnet. This imaging configuration provided a radiographic picture of the electron beam profile and its perturbation from the interaction fields.

By employing distinct object-to-detector distances for the X-ray and electron-beam imaging setups, the multi-messenger configuration provides a dual perspective of the interaction with different magnifications. The electron radiography setup allowed for a broader field of view of the beam profile, complementing the higher-resolution perspective offered by the X-ray imaging configuration.


\subsection{\label{subsec:MethodsEbeamCharacterize} Electron beam characterization}

The relativistic electron beam produced by the wakefield accelerator was characterized using a magnetic spectrometer~\cite{tsai2018control}, comprising a  $1.5 - 3$~T  electromagnet and a downstream  $1$~T  permanent dipole magnet along the beamline. The magnetic fields dispersed the electrons based on their momenta, projecting their trajectories onto a series of Lanex scintillating screens, from which the electron-induced fluorescence was imaged onto an array of 12-bit CCD cameras.

Characterization of the beam \hl{is} determined by mapping the electron positions on the screens to particle tracking simulations through the experimentally measured magnetic field. This process allows for precise \hl{beam} energy, divergence, and charge \hl{calibration} \hl{as described in Ref.
}~\cite{nakamura2011electron}.

\begin{figure*}
  \centering
  \includegraphics*[width=\linewidth]{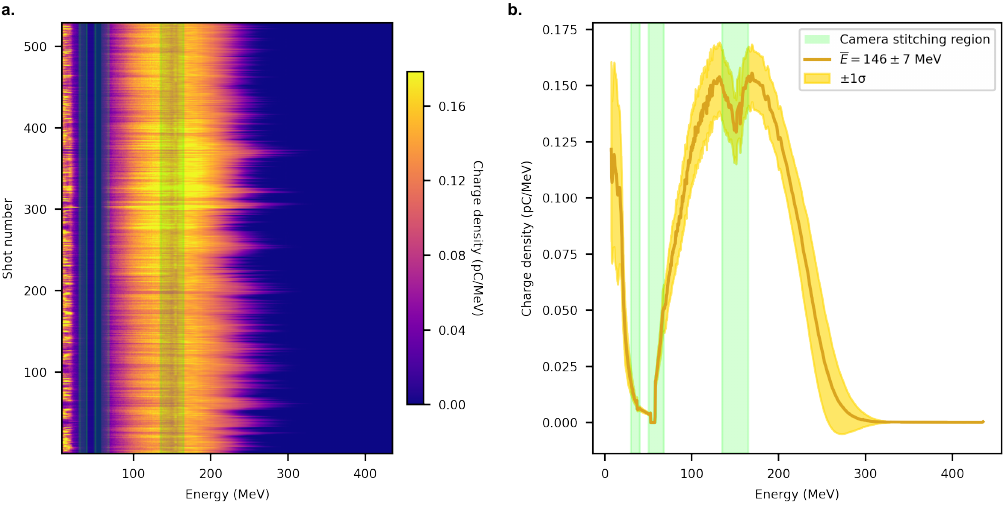}
  \caption{Characterization of electron beam probe using magnetic spectrometer. \textbf{a} \hl{529} shots were taken continuously measuring the mean momentum, charge, and FWHM divergence angle. \textbf{b} Electron beam spectrum including mean spectrum and energy spread.}
  \label{fig:ebeam-spectrum}
\end{figure*}

The mean energy of the electron beam for the experiment was \hl{$146\pm7$}~MeV as shown in Fig~\ref{fig:ebeam-spectrum}, with a pointing divergence of  $7.8 \pm 1$~mrad  in the $\mathbf{x}$-direction and $3.4 \pm 0.8$~mrad in the $\mathbf{y}$-direction. The mean charge of the beam was \hl{$24 \pm 4$~pC}. 

\hl{For the specific analysis of field recovery from electron radiographs, a representative beam energy of $E_{0} = 44$~MeV was selected. This lower energy reflects operating conditions where the beam was intentionally detuned to increase divergence and maximize field sensitivity across the imaging region. Additionally, this choice allows chromatic deflection effects to be better accounted for when measuring field magnitudes}.


\subsection{\label{subsec:MethodsGasJet} Gas target}

The laser plasma accelerator employed ionization injection ~\cite{McGuffey2010Ionization} to generate a relativistic electron beam by focusing the high-power laser pulse into a  $3 \, \text{mm}$  mixed gas jet ~\cite{zhou2021effect} composed of $99.5\%$  Helium and  $0.5\%$  Nitrogen. The supersonic gas jet utilized a fast solenoid valve (Parker Pulse Valve) synchronized and triggered alongside the high-power beamline. A voltage-controlled regulator allowed precise tuning of the gas jet’s backing pressure, enabling adjustment of the gas plume density within the range  $n_0 \in [2.1, 2.8] \times 10^{18} \, \text{cm}^{-3}$. To characterize the laser-plasma interaction, a Mach-Zehnder interferometer was utilized, incorporating a frequency-doubled probe beam oriented perpendicular to both the gas jet and the high-power laser line. This configuration enabled synchronized diagnostics of the plasma density profile~\cite{plateau2010wavefront}.



\subsection{\label{subsec:MethodsWater} Water target}

A cylindrical liquid water jet flowing in vacuum was employed as the target for the experiment, drawing inspiration from existing systems such as bulk liquid targets used in time-of-flight mass spectrometers \cite{charvat2004new} and thin liquid sheets and cryogenic jets developed for ion acceleration \cite{koralek2018generation, kim2016development}.

The water was delivered to the chamber by a High-performance liquid chromatoguraphy (HPLC) pump connected to 1/16" OD stainless steel tubing (0.03" ID). The jet nozzle consisted of cleaved Polymicro (Molex) capillary tubing with $30$~$\mu$m ID and $360$~$\mu$m OD. The water was delivered at a flow rate kept between $1-2$~ml min$^{-1}$, maintained constant by the HPLC pump. 

The water collector assembly included a custom-made, conical-shaped, copper piece with a top aperture of $d \approx 500$ $\mu$m, designed to capture the water stream. To prevent ice formation during the alignment procedure, the collector was heated by miniature cartridge heaters (Thorlabs, 15W) embedded in its copper head. The collector assembly, connected to a reservoir via standard DN16CF flanges, was kept at vapor pressures ($\sim 20 Torr)$ due to evaporation. 

The alignment of the water jet nozzle with the collector was achieved using picomotors for small adjustments along the $\mathbf{x}, \mathbf{y},$ and $\mathbf{z}$ directions. Two orthogonal cameras positioned outside the vacuum chamber monitored the procedure, and stable operation was achieved once the nozzle and collector were precisely aligned after pump-down. 

After alignment, chamber pressures as low as $10^{-5}$~Torr could be achieved using  $>1000$~L s$^{-1}$ turbo pumps and a $L N_{2}$ cold trap. In the absence of the cold trap, a pressure of $10^{-4} \, \text{Torr}$ could be maintained in the chamber. The stable water jet was subsequently aligned with the long pulse laser beam using translation stages that supported the entire nozzle-collector assembly.


\subsection{\label{subsec:MethodsFLASH} FLASH simulations}

 Three-dimensional (3D) radiation hydrodynamic simulations were performed using FLASH~\cite{fryxell2000flash} to predict the evolution of the laser-plasma interaction. FLASH is a multi-dimensional, radiation hydrodynamic simulation code based on the Eulerian approach. It includes dynamic block adaptive meshing, treats multiple materials, includes electron heat conduction and related physics, transports radiation via multigroup diffusion, and deposits laser energy by tracing rays in 3D. Although FLASH simulations accurately model the fluid dynamics, they do not include electric ($E$) and magnetic ($B$) fields nor treat hot electron populations.

The simulations utilized a 3D Cartesian coordinate system consisting of a cylindrical plasma target with a radius of $r_{0} = 15~\mu\text{m}$ and mass density $\rho_{\text{t}} = 1~\text{g cm}^{-3}$. A surrounding chamber plasma where $r > r_{0}$ was initialized with a much lower initial mass density $\rho_{\text{c}}$. Two different density profiles for the chamber plasma were explored: 1) a standard target with constant $\rho_c = 10^{-6}~\text{g} {cm}^{-3}$ throughout the domain, and 2) an evaporative target so that $\rho_c = \rho_{0} \left( r_{0}/ \sqrt{\mathbf{r}^{2}} \right)$ \hl{with $\rho_0 = 10^{-3}~\text{g cm}^{-3}$} following a decaying distribution as $r$ increases. 

Both the target and chamber plasmas were modeled with the same effective atomic number $Z_{\text{eff, t}} = Z_{\text{eff, c}} = 3.33$ and average atomic mass $A_{\text{eff, t}} = A_{\text{eff, c}} = 6.0$. The plasmas were similarly initialized at room temperature $T_{\text{t},0} = T_{\text{c},0} = 290~\text{K}$, and the multi-group flux-limited diffusion coefficient was chosen to be $f=0.17$, \hl{considering our laser parameters and Ref.}~\cite{chen2023effect}. The equation of state for both plasmas was calculated using the PrOpacEOS code.

FLASH utilizes geometric optics for laser energy deposition in the form of a ray-tracing model. A beam is represented by a number of rays whose paths are traced through the simulation domain, with their trajectories updated according to the local refractive index of each cell. The shock-driver laser pulse in the simulations was modeled with $\lambda = 0.8~\mu\text{m}$ and using a truncated Gaussian profile for its intensity with $\text{FWHM} = 40~\mu\text{m}$. The laser power deposited into each cell is calculated based on inverse bremsstrahlung, which depends on the local electron number density gradients and the local temperature gradients. The laser pulse was defined in sections using a piecewise linear function, where each section is associated with a time-power pair. These pairs ensure that the total energy in the pulse ($1$~J) is delivered over the total time window ($220$~ps).

\subsection{\label{subsec:MethodsImageProcessing} Image processing}

\subsubsection{\label{subsubsec:ImageFourier} Fourier mask}

We applied a custom filter in the Fourier domain to process the experimental data, reducing undesired high spatial frequencies and thereby increasing the signal-to-noise ratio. The method involves constructing a mask that selectively filters spatial frequencies based on a combination of radial and elliptical criteria. 

Given an input image $I(x, y)$ of dimensions $N_x \times N_y$, the two-dimensional Fourier transform  $\tilde{I}(k_x, k_y) = \mathcal{F}\{ I(x, y)\}$ is computed first, where  $k_x$  and $k_y$  represent the spatial frequency components in the  $x$ and  $y$ directions, respectively. 

Next, a radial frequency component $K = \left(k_x^2 + k_y^2\right)^{1/2}$ and an elliptical frequency component $K_{\text{el}} = \left(a^2 k_x^2 + b^2 k_y^2 \right)^{1/2}$ are defined. The final frequency-space mask $M(k_x, k_y)$ is then constructed by incorporating these two terms as:
\begin{equation}
\begin{split}
M(k_x, k_y) = &\left( 1 - \exp\left[ - \left( \frac{K}{K_{\text{max}}} \cdot 100 \right)^b \right] \right) \\ 
&\times \left( \frac{ | \sin(\pi K_{\text{el}}) | }{ \pi K_{\text{el}}} \right)^{N}
\end{split}
\end{equation}
where $K_{\text{max}} = \pi$  is the maximum spatial frequency, and  $N$  is a parameter that controls the sharpness of the mask. To avoid singularities when $K_{\text{el}} = 0$, we set the corresponding values in the mask to $1$. Moreover, prior to applying the frequency-space mask the image is rotated by an angle $\theta$ to align it with the \hl{vertical} orientation using bilinear interpolation. \hl{In this work, we selected $a = 0.5$, $b = 2.0$, and $N = 16$ to optimize the quality of the processed experimental images.}

The constructed mask in frequency-domain is then applied to the image as $\tilde{I}_{\text{masked}}(k_x, k_y) = \tilde{I}(k_x, k_y) \cdot M(k_x, k_y)$, and the inverse Fourier transform is used to recover the \hl{masked} image in real space:
\begin{equation}
I_{\text{masked}}(x, y) = \mathcal{F}^{-1}\{ \tilde{I}_{\text{masked}}(k_x, k_y) \}
\end{equation}

\subsubsection{\label{subsubsec:ImageFresnel} Fresnel-Kirchoff algorithm}

To model the expected image pattern from a phase-contrast imaging system, we follow a common approach outlined by Born and Wolf ~\cite{born2013principles} and applied in past work ~\cite{arfelli1998low, olivo2006experimental}, which consists of solving the Fresnel-Kirchoff integrals numerically to obtain the expected complex wave-field distribution $\tilde{u}$ after propagation from an initial point at the source, to a point $P$ in the detector plane. In this context, the real ``pure" pattern $I$ is given by 
\begin{equation}
    I = I_{0} \mid \Tilde{u}(P) \mid^{2}\;,
\end{equation}
where $I_{0}$ is the incoming intensity distribution at the sample. 

Assuming that the effective Fresnel number is larger than unity, the complex-valued wave-field distribution can be simplified utilizing the paraxial approximation. This approximation yields a final expression for the complex-valued wave-field distribution in Fourier space \hl{as given by} ~\cite{barbato2019quantitative},
\begin{equation}
\begin{split}
    \tilde{U}(u,v) = &M^{2} T(Mu, Mv) \times \\
    &\text{exp}\left[ -\pi i \lambda R_{1} M (u^{2} + v^{2}) \right] \times \\
    &\text{exp}\left[ 2 \pi i (R_{1}/R_{0}) (x_{0} u + y_{0} v) \right]\;,
    \label{eq:fourier-wave}
\end{split}
\end{equation}
where $u$ and $v$ are the transverse spatial frequencies corresponding to $x$ and $y$ coordinates, respectively. Equation~\ref{eq:fourier-wave} can be solved using a fast Fourier transform (FFT) algorithm to obtain the pure pattern $I$, which can then be convolved with the source size $\sigma$ to obtain the real phase-contrast pattern in the detector. Here $T(Mu, Mv)$ is the Fourier transform of the object transfer function $t(P) = \text{exp}\left[i \phi \right]$, where the phase induced $\phi$ is complex-valued and depends on the material properties as
\begin{equation}
    t(P) = \text{exp}\left[ i \left( i D(P) - B(P) \right) \right]
    \label{eq:transfer-matrix}
\end{equation}
For X-rays crossing a sample the index of refraction is less than unity and has the form $n = 1 - \delta - i \beta$. The phase map is then calculated using Eq.~\ref{eq:transfer-matrix} and the following projected distributions obtained from radiation hydrodynamic simulations,
\begin{equation}
    D(P) = \frac{- 2 \pi}{\lambda} \int \delta(x,y,z) dr
\end{equation}
and
\begin{equation}
    B(P)  = \frac{- 2 \pi}{\lambda} \int \beta(x,y,z) dr\;.
\end{equation}
where  $\delta$  and  $\beta$  are the real and imaginary components of the refractive index and are expressed as,
\begin{equation}
    \delta = \frac{r_{e} N_{A} \lambda^{2} \rho}{2 \pi} \sum_{j} \frac{w_{j} \left[ Z_{j} + f'_{j} \right]}{A_{j}}
\end{equation}
and,
\begin{equation}
    \beta = \frac{\nu}{\omega_{p}} \delta
\end{equation}
where $r_{e}$ is the classical electron radius, $N_{A}$ is the Avogadro number, $\rho$ is the mass density, $A_{j}$ are the atomic number and the atomic weight of the $j$-th element of molecule, $f'_{j}$ is the real part of the dispersion factor, $\omega_{p}$ is the density-dependent plasma frequency, and $\nu$ is the collisions rate factor. 

Finally, to account for the polychromaticity of the source, the resulting image can be weighted and summed according to the spectrum distribution of the source
\begin{equation}
    I_{poly} = \sum_{\lambda} w(E) I_{mono} (\lambda)\;,
\end{equation}
where $w(E)$ is the energy-dependent weighting factor obtained from the source spectrum.

\subsubsection{\label{subsubsec:ImageMeijering} Meijering filter}

The Meijering filter is \hl{an image processing} technique to accurately quantify and segment neurite-like traces in fluorescence microscopy images, as described by Meijering \textit{et al.}~\cite{meijering2004design}. In this work the algorithm is adapted for the detection of shocked traces in the sample, which deviate from the nominal un-shocked image. 

For the implementation of the algorithm, the modified second-order derivatives of the image are calculated by convolving it with the second-order derivatives of the Gaussian kernel. Specifically, if $f$ mathematically represents the image and $G$ denotes the normalized Gaussian kernel, the second-order derivative of the image at position $\vec{x} = (x,y)$ can be computed as,
\begin{equation}
f_{ij}(\vec{x}) = \left( f * G_{ij} \right)(\vec{x}) 
\end{equation}
where
\begin{equation}
G_{ij}(\vec{x}) = \left( \frac{\partial^{2}}{\partial_i \partial_j} G \right)(\vec{x})
\end{equation}
and the derivative directions $i$ and $j$ can be either $x$ or $y$. 

Next, the eigenvectors and eigenvalues are computed in the Meijering algorithm by using a modified second-derivative matrix given by,
\begin{equation}
    H'_{f}(\vec{x}) = \begin{bmatrix}
f_{xx}(\vec{x}) + \alpha f_{yy}(\vec{x}) & \left( 1 - \alpha \right) f_{xy}(\vec{x}) \\
\left( 1 - \alpha \right) f_{xy}(\vec{x}) & f_{yy}(\vec{x}) + \alpha f_{xx}(\vec{x}) 
\end{bmatrix}
\end{equation}
where $\alpha$ is a parameter which needs to be optimized. The normalized eigenvectors $\text{v}_{\text{i}}(\vec{x})$ and their corresponding eigenvalues $\lambda_{\text{i}}(\vec{x})$ of the standard Hessian second-derivative matrix are then obtained as,
\begin{equation}
\begin{cases} 
      \text{v}'_{1}(\vec{x}) = v_{1}(\vec{x}) \\
      \text{v}'_{2}(\vec{x}) = v_{2}(\vec{x})  \\ 
\end{cases}
\end{equation}
and
\begin{equation}
\begin{cases} 
      \lambda'_{1}(\vec{x}) = \lambda_{1}(\vec{x}) + \alpha \lambda_{2}(\vec{x}) \\
      \lambda'_{2}(\vec{x}) = \lambda_{2}(\vec{x}) + \alpha \lambda_{1}(\vec{x}) \\ 
\end{cases}
\end{equation}
In this way, the algorithm calculates the eigenvectors of the Hessian to determine the similarity of an image region to the shocked traces in the sample. \hl{The Meijering filter was applied to both experimental and simulation images using its default parameters as given by scikit-image filters package.}

\newpage 

\subsubsection{\label{subsubsec:ImageFields} Field recovery method}

\begin{figure*}
  \centering
  \includegraphics*[width=0.7\linewidth]{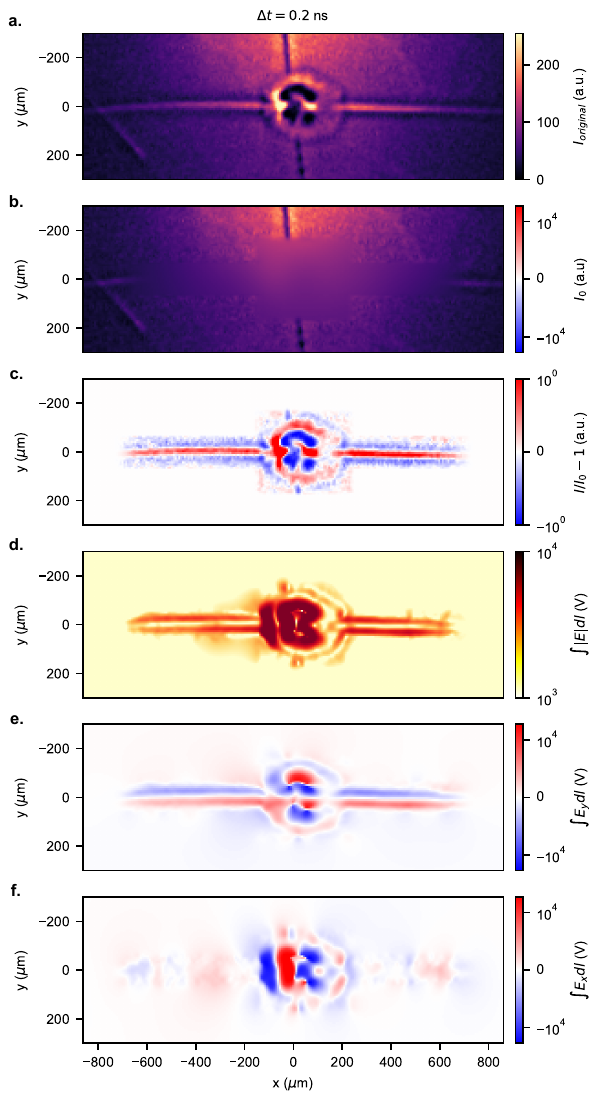}
  \caption{Electric field recovery process from experimental electron beam radiographic images. \hl{\textbf{a}} Raw normalized image $I$ from electron beam probe on phosphorus screen. \hl{\textbf{b}} Inferred unperturbed beam $I_{0}$ from blurring and masking. \hl{\textbf{c}} \textit{Difference} image ($I/I_{0} - 1$) where red indicates extra electrons and blue indicates fewer electrons. \hl{\textbf{d}} Recovered projected electric field magnitude $| E |$ along electron beam probe direction. \hl{\textbf{e}} Vertical component of projected electric field $E_{y}$. \hl{\textbf{f}} Horizontal component of projected electric field $E_{x}$. }
  \label{fig:ED-field-recovery}
\end{figure*}

Following the technique described by Kugland \textit{et al.}~\cite{kugland2012invited}, the field recovery method begins by assuming that the perturbed object is located at a distance $R_{0}$ from the particle (electron) source, with the detection screen situated at a distance $R_{1}$ from the object. Generally, $R_{1} \gg R_{0}$, and $R_{0} \gg a$, where $a$ is a characteristic spatial scale of the field in question. In a Cartesian geometry, the ideal image $I_{0}$ at the object plane for an undisturbed electron beam (see Fig.~\ref{fig:ED-field-recovery}) is described by the coordinates $(x_{0}, y_{0})$, while the real image $I$ in the detector is described by the coordinates $(x,y)$, such that $x = x_{0}\times M$ and $y=y_{0} \times M$, where $M=(1+R_{1}/R_{0})$ is the geometric magnification. 

After the electrons traverse the electromagnetic fields they are deflected by angles $\alpha_{x}$ and $\alpha_{y}$. The coordinates of the real image, as described by the electron trajectories can be approximated, when $\alpha_x,y$ are small, by
\begin{equation}
    x = x_{0} + \frac{R_{1}}{R_{0}} x_{0} + \alpha_{x} R_{1}\;,
\end{equation}

\begin{equation}
    y = y_{0} + \frac{R_{1}}{R_{0}} y_{0} + \alpha_{y} R_{1}\;,
\end{equation}
Therefore, the objective of the algorithm is to determine the deflection angles $\alpha_{x}$ and $\alpha_{y}$, as they can be related to the path-integrated electric or magnetic fields present at the interaction region by the following expressions,
\begin{equation}
\begin{split}
    &\alpha_{x}^{+} = \frac{q}{\gamma m_{e} v^{2}_{z}} \int E_{x} dz \\ &\alpha_{x}^{-} = \frac{q}{\gamma m_{e} v_{z}} \int B_{y} dz\;,
\end{split}
\end{equation}
and
\begin{equation}
\begin{split}
    &\alpha_{y}^{+} = \frac{q}{\gamma m_{e} v^{2}_{z}} \int E_{y} dz \\ &\alpha_{y}^{-} = \frac{q}{\gamma m_{e} v_{z}} \int B_{x} dz\;.
\end{split}
\end{equation}
Finally, the disturbed beam at the detector plane $I(x,y)$ can be then obtained by the following relation,  
\begin{equation}
    I(x,y) = \frac{I_{0}(x_{0},y_{0})}{ \left\|{\frac{\partial(x,y)}{\partial(x_{0},y_{0})}} \right\|}\;,
\end{equation}
Here, $\mid \partial(x,y) / \partial(x_{0},y_{0}) \mid$ is the absolute value of the determinant of the Jacobian matrix that relates the object and image planes and can is described by
\begin{equation}
\begin{split}
    \lvert \frac{\partial(x,y)}{\partial(x_{0},y_{0})} \rvert =  & \lvert \left(1 + \frac{L}{l} + \frac{\partial \alpha_{x}}{\partial x_{0}} L \right) \\
    &\times \left( 1 + \frac{L}{l} + \frac{\partial \alpha_{y}}{\partial y_{0}} L \right) \\
    &- L^{2} \frac{\partial \alpha_{x}}{\partial y_{0}} \frac{\partial \alpha_{y}}{\partial x_{0}} \rvert  \;.
\end{split}
\end{equation}
To solve this equation, we assume that $\pmb{\alpha} = [\alpha_x,\alpha_y,0]$ is small and that the object-to-image coordinates follow a linear mapping. Additionally, all electron trajectories are assumed to be rotation-less such that $\nabla \times \pmb{\alpha} = 0$. For this regime, the deflection angles can be obtained from a potential field $\phi$ so that
\begin{equation}
    \pmb{\alpha} = \nabla {\Phi}\;.
\end{equation}
For convenience, we introduce the following normalization: $\tilde{x}=x/w_{x}$, $\tilde{y}=y/w_{y}$, $\tilde{\alpha_i} = \alpha_i L / w_i$, $\tilde{\pmb{\alpha}} = \tilde{\nabla} \tilde{{\Phi}}$, and $\tilde{I_{0}} = I_{0}/M^{2}$ where $w_i$ is width of the beam at the image plane. The Jacobian, described by Kugland \textit{et. al}, can then be rewritten as,
\begin{equation}
\begin{split}
    &\frac{1}{M^{2}} \left\| \frac{\partial(x,y)}{\partial(x_{0},y_{0})} \right\| = \\
    &\left\| 1 + \tilde{\nabla}^{2} \tilde{{\Phi}} + \frac{\partial^{2} \tilde{{\Phi}}}{\partial \tilde{x}^{2}} \frac{\partial^{2} \tilde{{\Phi}}}{\partial \tilde{y}^{2}} - \left( \frac{\partial^{2} \tilde{{\Phi}}}{\partial \tilde{x} \partial \tilde{y}} \right)^{2} \right\| \;.
\end{split}
\end{equation}
This leads to the final equation, which is solved numerically,
\begin{equation}
    \tilde{\nabla}^{2} \tilde{{\Phi}} = \frac{\tilde{I}_{0} + \epsilon}{I + \epsilon} - 1 - \frac{\partial^{2}\tilde{{\Phi}}}{\partial \tilde{x}^{2}} \frac{\partial^{2}\tilde{{\Phi}}}{\partial \tilde{y}^{2}} + \left( \frac{\partial^{2} \tilde{{\Phi}}}{\partial \tilde{x} \partial \tilde{y}} \right)^{2}\;.
\end{equation}
\hl{Under strong-field conditions, or in the presence of low-energy electrons, large deflection angles $\alpha_{x,y}$ can lead to overlapping beam trajectories, and the formation of caustics in the radiograph. This affects the one-to-one mapping between the source and detector coordinates, making the relationship between $x_0$ and $x$ nonlinear and the correspondence between $I_0$ and $I$ non-unique. While the broad energy spectrum of the electron beam can further introduce chromatic effects in the deflection response, these should not be large as described in Ref.}~\cite{zhang2016capturing}.\hl{ Key features of the field topology, its evolution in time, and approximate lower bounds on field strength may still be robustly extracted using this method.}





\backmatter





\section*{Data Availability} 
The data supporting the findings of this study are included in the main text Results and Extended Data sections.
Source data has been deposited in the Michigan Deep Blue Data Repository. Additional data is available from the corresponding author upon request.

\section*{Code Availability} 
Code for producing figures in the main text Results and Extended Data sections, including image processing and simulation scripts, have been deposited in the Michigan Deep Blue Data Repository. 
Additional code for processing raw experiment data is available from the corresponding author upon request.

\newpage

\newpage

\bibliography{sn-bibliography}




\section*{Acknowledgments}

The work was supported by the U.S. Department of Energy (US DOE) Office of Science Fusion Energy Sciences Grant No. DE-SC0020237,  Contract No. DE-AC02-05CH11231 the LaserNetUS initiative at the Berkeley Lab Laser Accelerator (BELLA) Center, and partially supported under Contract No. FWP-100884: LaserNetUS Management. The work was also supported by US DOE National Nuclear Security Administration (NNSA) Center of Excellence under Cooperative Agreement No. DE-NA0003869,  the European Research Council (ERC) under the European Union’s Horizon 2020 research and innovation programme with grant agreement No. 682399, and  Lawrence Livermore National Laboratory under subcontract No. B645096. Lawrence Livermore National Laboratory is operated by Lawrence Livermore National Security, LLC, for the US DOE NNSA under Contract No. DE-AC52-07NA27344.


\section*{Competing Interests} 

The authors declare no competing interests. 

\newpage

\section*{Author Contributions}


M.D. Balcazar: 
Conceptualization, Data curation, Formal analysis, Investigation, Methodology, Software, Visualization, Writing – original draft, Writing – review \& editing.
H-E Tsai: 
Data curation, Investigation, Methodology, Project Administration, Supervision, Writing – review \& editing.
T. Ostermayr:
Investigation, Methodology, Software, Writing – review \& editing.
P.T. Campbell: 
Investigation, Methodology, Software, Validation, Writing – review \& editing. 
F. Albert: 
Writing – review \& editing.
Q. Chen: 
Investigation, Writing – review \& editing.
C. Colgan: 
Investigation, Writing – review \& editing.
G. Dyer: 
Resources, Writing – review \& editing.
Z. Eisentraut: 
Investigation, Methodology, Writing – review \& editing.
E. Esarey: 
Writing – review \& editing.
C.G.R. Geddes: 
Investigation, Methodology, Resources, Project administration, Writing – review \& editing.
E.S. Grace: 
Investigation, Writing – review \& editing.
B. Greenwood: 
Investigation, Writing – review \& editing.
A. Gonsalves: 
Investigation, Methodology, Software, Writing – review \& editing.
S. Hakimi: 
Investigation, Writing – review \& editing.
R. Jacob: 
Data curation, Investigation, Methodology, Writing – review \& editing.
B. Kettle: 
Investigation, Writing – review \& editing.
P. King: 
Investigation, Methodology, Writing – review \& editing.
K. Krushelnick: 
Writing – review \& editing.
N. Lemos: 
Writing – review \& editing.
E. Los: 
Investigation, Writing – review \& editing.
Y. Ma: 
Investigation, Writing – review \& editing.
S.P.D. Mangles: 
Writing – review \& editing.
J. Nees: 
Conceptualization, Investigation, Methodology, Writing – review \& editing.
I.M. Pagano: 
Methodology, Writing – review \& editing.
C. Schroeder: 
Writing – review \& editing.
R.A. Simpson: 
Investigation, Writing – review \& editing.
A.G.R. Thomas: 
Conceptualization, Formal Analysis, Funding acquisition, Investigation, Methodology, Project administration, Resources, Software, Supervision, Writing – review \& editing.
M. Trantham: 
Methodology, Software, Validation, Writing – review \& editing.
J. Van Tilborg: 
Investigation, Project administration, Resources, Writing – review \& editing.
A. Vazquez: 
Investigation, Writing – review \& editing.
C.C. Kuranz: 
Conceptualization, Funding acquisition, Investigation, Methodology, Project administration, Resources, Writing – review \& editing.






\section*{Extended Data}\label{sec:ExtendedData}


\begin{figure*}
  \centering
  \includegraphics*[width=0.5\linewidth]{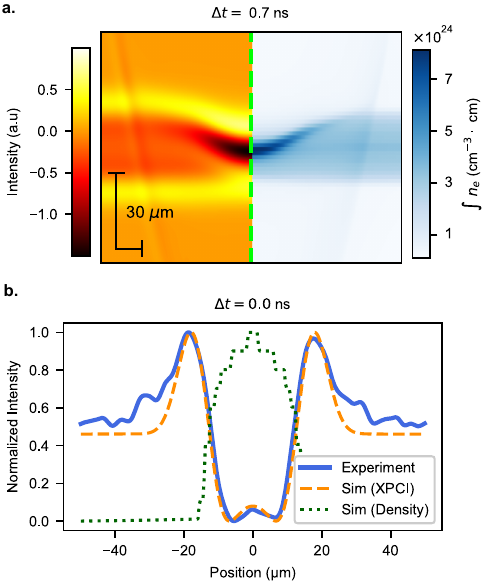}
  \caption{ Comparison of 3D FLASH simulations with synthetic phase-contrast X-ray imaging, accompanied by calibration relative to experimental measurements. \hl{\textbf{a} Illustration of mapping between FLASH projected density distributions to synthetic phase-contrast X-ray images with Fresnel-Kirchoff algorithm. \textbf{b} Comparison of center lineouts between density simulation, synthetic phase-contrast image and experimental data. This illustrate calibration for comparison between simulation and experiment.}}
  \label{fig:FLASH-XPCI}
\end{figure*}


\end{document}